\documentclass[amsmath, amsfonts, showpacs, superscriptaddress, twocolumn, prx]{revtex4}
\usepackage{graphicx}
\usepackage{epsfig}
\usepackage{bm}
\usepackage{dcolumn}
\usepackage{amsmath}
\usepackage{amssymb}
\usepackage{color}
\usepackage{stackrel}
\usepackage{accents}
\usepackage{latexsym}
\usepackage{verbatim}
\usepackage{hyperref}

\newcommand{\beg}{\begin{equation}}
	\newcommand{\en}{\end{equation}}

\newcommand{\bp}{\mathbf p}

\newcommand{\br}{\mathbf r}

\newcommand{\bR}{\mathbf R}
\newcommand{\dg}{^\dagger}
\newcommand{\up}{\uparrow}
\newcommand{\dn}{\downarrow}

\begin{document}
	
	\title{Quasiclassical circuit-theory of contiguous disordered multiband superconductors}
	
	\author{Ammar A. Kirmani}
	\affiliation{Department of Physics, Kent State University, Kent, Ohio 44242, USA}
	
	\author{Maxim Dzero}
	\affiliation{Department of Physics, Kent State University, Kent, Ohio 44242, USA}

	\author{Alex Levchenko}
	\affiliation{Department of Physics, University of Wisconsin-Madison, Madison, Wisconsin 53706, USA}
	\affiliation{Max Planck Institute for the Physics of Complex Systems, N\"othnitzer str. 38, 01187 Dresden, Germany} 
	
	\vspace{10pt}
	
	\begin{abstract} 
		We consider a general problem of a Josephson contact between two multiband superconductors with coexisting superconducting and magnetic phases. As a particular example, we use the quasiclassical theory of superconductivity to study the properties of a Josephson contact between two disordered $s^{\pm}$-wave superconductors allowing for the coexistence between superconductivity and  spin-density-wave orders. The intra- and inter-band disorder-induced scattering is treated within the self-consistent Born approximation. We calculate the spatial profile of the corresponding order parameters on both sides of the interface with a finite reflection coefficient and use our results to evaluate the local density of states at the interface as well as critical supercurrent through the junction as a function of phase or applied voltage. Our methods are particularly well suited for describing spatially inhomogeneous states of iron-based superconductors where controlled structural disorder can be created by an electron irradiation. We reveal the connection between our theory and the circuit-theory of Andreev reflection and extend it to superconducting junctions of arbitrary nature. Lastly, we outline directions for further developments in the context of proximity circuits of correlated electron systems.
	\end{abstract}
	
	
	\date{\today}
	
	\maketitle
	
	\section{Introduction}
	
	In many practical cases superconductivity occurs in the form of a spatially inhomogeneous state \cite{TheoryOfMetals,Svidzinsky}. This can be triggered intrinsically due to thermodynamic reasons or created extrinsically by forming contacts of superconductors with other materials. The fundamental example of the first kind of inhomogeneity is given by the Abrikosov vortex state, which brings about the spatial modulation of the order parameter \cite{TheoryOfMetals,IoffeReview}. Josephson junction is the example of the other kind, where inhomogeneity is created near the contact area when two superconductors are brought into proximity via a tunnel barrier or other type of the weak link \cite{Josephson-RMP,Golubov-RMP}. 
	In both of these cases, and many other physical situations, the spatial inhomogeneity extends over the length scale of superconducting coherence length that is large as compared to electron Fermi wavelength. Under this condition, the semiclassical theory of superconductivity based on Eilenberger \cite{Eilenberger1968} and Usadel \cite{Usadel1970} equations become applicable. These two methods were developed to treat relatively clean and strongly disordered superconductors respectively. The solutions of the Eilenberger and Usadel equations relate the observable properties of a superconducting structure, for example critical current, to microscopic characteristics of the materials forming the junction and its geometry.  
	
	An alternative method to describe mesoscopic superconducting structures is based on the random matrix and scattering matrix theories  \cite{Beenakker-RMP}. In this approach, all microscopic details are condensed into symmetry properties of the scattering matrix representing a disordered region of the junction which is typically parametrized by a set of transmission eigenvalues. An observable of interest is then expressed in terms of these transmissions similar to the Landauer-Buttiker transport theory. This phenomenology is more straightforward and intuitive than semiclassical kinetic theory, but it is more restrictive in terms of conditions when it applies. Yet there is a parameter range when both methods work, however, the connection between them is not immediately obvious. 
	
	This link has been provided by the circuit-theory of Andreev reflection developed originally by Nazarov \cite{Nazarov1994}, later reviewed and extended by several authors \cite{Beenakker-RMP,Nazarov,Argaman}. Circuit theory can be formulated as the finite set of rules for connectors and nodes of a given superconducting devices, analogous in spirit to Kirchhoff's rules. It also gives a prescription to deal with boundary conditions and in particular average over the transmission eigenvalues \cite{Dorokhov,Schep}, which are in general random for a disordered or chaotic junction between superconductors.        
	In recent years, we witnessed the emergence of novel classes of multiband unconventional superconductors, primarily the large family of iron-pnictides (see reviews \cite{Matsuda_review2014,Chubukov} and references therein). Semiclassical methods of superconductivity were successfully applied to describe their properties including the proximity and Josephson effects \cite{Moor,Apostolov,Vakaryuk,Lin,Stanev,Koshelev,Golubov,Berg,Yerin,Ota-1,Ota-2,Chen,Tsai,Linder,Nagaosa} but circuit-theory has not been derived for these systems. The motivation for this work is to put forward a detailed theory of superconducting contacts, where the material constituents forming the junction harbor complex superconducting phases and competing order parameters.  
	
	This paper is organized as follows. In Section II, we formulate the simplest two-band model that allows for the coexistence of superconducting (SC) and spin-density-wave (SDW) orders, and derive the Eilenberger equations, which form the technical basis for our work. In the Section III, we employ the method developed by Yip \cite{Yip1997} to solve the quasiclassical equations and, at the same time, satisfy full nonlinear boundary conditions derived by Zaitsev \cite{Zaitsev1984}. In addition, we have arbitrary transparencies and shapes of potential barriers forming the constriction. We demonstrate, that the special auxiliary decomposition of nonlinear constraints naturally leads to the circuit-theory boundary conditions as elaborated by Nazarov \cite{Nazarov}. In Section IV, we present the results for the numerical solution for the spatially dependence of the superconducting order parameter, local density of states at the interface and Josephson current. In Section V, we briefly review several universal examples of the Josephson effect in mesoscopic superconductor-normal-superconductor (SNS) structures and related devices with insulating barriers and micro-constrictions. We discuss how circuit-theory captures in a unified fashion multiple results for the Josephson current-phase relationships that were previously known from the separate semiclassical calculations and extend that to the case of proximity junctions of correlated electrons. Section VI is devoted to the discussion of the results and outlook for further developments. Lastly, in Appendices A, B \& C, we provide the details on the derivations of the expressions that we used to obtain the solution of the Eilenberger equations. %\tcr{some typos fixed and commas added A.K.}

%###############################################################################################################
	
	\section{Formulation of the problem}
	In what follows, we introduce the model Hamiltonian and write down the quasiclassical equations for the correlation functions which are used to determine the spatial profile of the superconducting and magnetic order parameters across the interface. 
	We will consider short junctions at an arbitrary transparency between two multiband superconductors. Generally, disorder in these systems induces both intra-band and inter-band scattering. In addition, the symmetry of the order parameter can be unconventional. As a guiding example, we will study a superconductor with the $s^{\pm}$ symmetry of the order parameter relevant for some classes of iron-pnictides. We will also consider a more complicated case, when the superconducting state coexists with another order such as spin-density wave. Under such circumstances, it will be impossible to write the Josephson current of such junctions just in terms of transmission eigenvalues. However, it is still possible to derive a closely related circuit-theory expression written in the form of semiclassical Green functions. 

%############################################################################################################### 
\begin{figure}[h]
	\includegraphics[width=8cm]{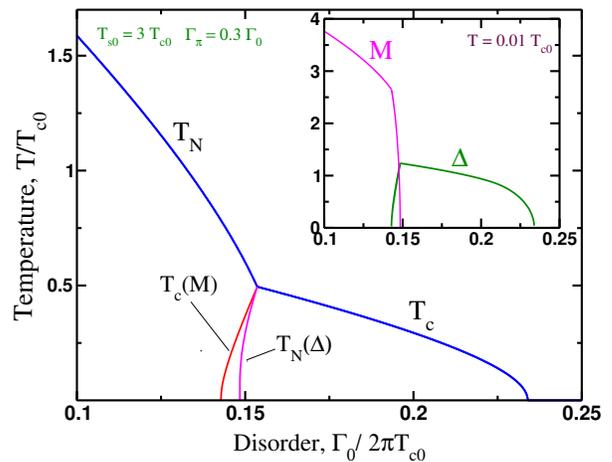}
	\caption{Phase diagram obtained by numerical solution of the mean-field equations (\ref{self-consistent}) and (\ref{MFEqs}) for a given 
		set of disorder scattering rates. The main panel shows doping dependence of the critical temperatures for magnetic ($T_N$) and superconducting ($T_c$) transitions. 
		The inset panel shows respective dependence of the order parameters $M$ and $\Delta$.}
	\label{Fig-PhaseDiagram}	
\end{figure}
%###############################################################################################################
	
	\subsection{Model Hamiltonian}
	Following the discussion in Refs. [\onlinecite{VC-PRB11,London2015}], we consider a model with two cylindrical Fermi surfaces. One Fermi surface has an electron-type (c) and the other one has a hole-type (f) excitations. We introduce the following eight-component spinor in momentum representation
	\beg\label{Spinor}
	\hat{\Phi}_\bp\dg=\left(\hat{\psi}_{\bp c}\dg, ~-i\hat{\sigma}_y\hat{\psi}_{-\bp c}^T, 
	~\hat{\psi}_{\bp h}\dg, ~-i\hat{\sigma}_y\hat{\psi}_{-\bp h}^T\right).
	\en
	Here $\hat{\sigma}_y$ is a Pauli matrix, $\hat{\psi}_{\bp a}\dg=(a_{\bp\up}\dg, ~a_{\bp\dn}\dg)$, $(a=c,f)$ and 
	$\hat{\psi}^T$ denotes the transpose of the operator. The form of (\ref{Spinor}) ensures the correction definition of the spin density operator
	at point $\br$:
	\beg\label{Spin}
	{\vec S}(\br)=\hat{\psi}_{c}\dg(\br){\vec \sigma}\hat{\psi}_{h}(\br)+
	\hat{\psi}_{h}\dg(\br){\vec \sigma}\hat{\psi}_{c}(\br).
	\en
	In this paper, however, we will limit our discussion to the $z$-component of the spin operator (\ref{Spin}), and as it turns out it will be more convenient to work with the following spinor 
	\beg\label{oldspinor}
	\hat{\Psi}_\bp\dg=\left(\hat{\psi}_{\bp c}\dg, ~\hat{\psi}_{-\bp c}^T,
	~\hat{\psi}_{\bp h}\dg, ~\hat{\psi}_{-\bp h}^T\right).
	\en
	
	The Hamiltonian for our problem can be written down using the mean-field approximation, 
	\beg\label{Eq1}
	\hat{H}=\hat{H}_0+\hat{H}_{\textrm{mf}}+\hat{H}_{\textrm{dis}}. 
	\en
	The noninteracting part $\hat{H}_0$ has the standard form pertinent to our choice of the basis spinor (\ref{oldspinor}):
	\beg\label{H0}
	\hat{H}_0=-\xi_\bp\hat{\tau}_3\hat{\rho}_3\hat{\sigma}_0, \quad \xi_\bp=\frac{p^2}{2m}-\mu,
	\en
	where  $\mu$ is a chemical potential and the mass anisotropy between hole- and electron-like bands was ignored. 
	The remaining mean-field part contains two terms which account for the superconducting pairing in $s^{\pm}$ symmetry channel with 
	the amplitude $\Delta$ and spin-density wave order parameter, ${\vec M}=M{\vec e}_z$:
	\beg\label{Hmf}
	\hat{H}_{\textrm{mf}}=-\Delta\hat{\tau}_3\hat{\rho}_2\hat{\sigma}_2+M\hat{\tau}_1\hat{\rho}_3\hat{\sigma}_3.
	\en
	In these expressions, we use the Pauli matrices ${\tau}_i$, ${\rho}_i$ and ${\sigma}_i$ $(i=1,2,3)$ defined in the subspace of band, Nambu and spin degrees of freedom correspondingly. 
	
	Lastly, the third term in (\ref{Eq1}) describes the effects of disorder induced by chemical substitution at lattice sites ${\mathbf R}_i$:
	\beg\label{disorder}
	\hat{H}_{\textrm{dis}}=\sum\limits_{i}\left[u_0\hat{\tau}_0\hat{\rho}_3\hat{\sigma}_0+u_\pi \hat{\tau}_1\hat{\rho}_3\hat{\sigma}_0\right]\delta(\br-{\mathbf R}_i).
	\en 
	The first term in the brackets accounts for the intraband scattering, while the second term describes the scattering between the bands. Having defined the Hamiltonian, next we outline the steps which lead to the equations for the quasiclassical Green's functions. 
	
	%###############################################################################################################
	\subsection{Eilenberger equation}
	For simplicity, let us first assume that disorder is the only source of spatial inhomogeneities. 
	To derive the equations for the quasi-classical correlation function, one starts with the Dyson equation for the single-particle Green's function in the imaginary time representation
	\beg\label{Grrp}
	\hat{G}(\br,\br';\tau)=-\langle\langle\hat{T}_\tau\{\hat{\Psi}(\br,\tau)\hat{\Psi}\dg(\br',0)\}\rangle\rangle_{\textrm{dis}}
	\en
	averaged over various disorder realizations:
	\beg\label{DysonEq}
	\left[i\omega_n-\hat{H}_\bp-\hat{\Sigma}(i\omega_n)\right]\hat{G}(\bp,i\omega_n)=\hat{1}
	\en 
	with $\omega_n=\pi T(2n+1)$ being the fermionic Matsubara frequency. Within the self-consistent Born approximation, the corresponding expression for the self-energy reads
	\beg\label{Sigma}
	\begin{split}
		\hat{\Sigma}(i\omega_n)&=\frac{\Gamma_0}{\pi\nu_F}\int\frac{d^2\bp}{(2\pi\hbar)^2}\hat{\tau}_0\hat{\rho}_3\hat{\sigma}_0
		\hat{G}(\bp,i\omega_n)\hat{\tau}_0\hat{\rho}_3\hat{\sigma}_0\\&+\frac{\Gamma_\pi}{\pi\nu_F}\int\frac{d^2\bp}{(2\pi\hbar)^2}\hat{\tau}_1\hat{\rho}_3\hat{\sigma}_0
		\hat{G}(\bp,i\omega_n)\hat{\tau}_1\hat{\rho}_3\hat{\sigma}_0,
	\end{split}
	\en
	where $\Gamma_{0,\pi}\propto \nu_F|u_{0,\pi}|^2$ are the corresponding disorder scattering rates and $\nu_F$ is the density of states at the Fermi level. 
	
	The quasiclassical Eilenberger function is defined according to
	\beg\label{calG}
	\hat{\cal G}(i\omega_n)=\frac{i}{\pi\nu_F}\int\frac{d^2\bp}{(2\pi)^2}\hat{\tau}_3\hat{\rho}_3\hat{\sigma}_0\hat{G}(\bp,i\omega_n).
	\en
	The equation for the function can be obtained from the Dyson equation above (\ref{DysonEq}) by eliminating the single particle spectrum, $\xi_\bp$. 
	
	In the spatially inhomogeneous case, which naturally arises in nonzero external magnetic field or in the presence of a contact between two superconductors, functions $\hat{G}(\bp,i\omega_n)$, $\hat{\cal G}(i\omega_n)$ as well as self-energy $\Sigma(i\omega_n)$ and the order parameters $\Delta$, $M$ will also depend on the 'center-of-mass' coordinate ${\mathbf R}=(\br+\br')/2$. Thus, function 
	$\hat{\cal G}({\mathbf R},i\omega_n,{\mathbf v})$ satisfies the following equation:
	\beg\label{Eq2}
	\begin{split}
		\left[i\omega_n\hat{\tau}_3\hat{\rho}_3\hat{\sigma}_0;\hat{\cal G}\right]&-i{\mathbf v}_F\cdot \partial_{{\mathbf R}} \hat{\cal G}-\left[\hat{H}_{\textrm{mf}}\hat{\tau}_3\hat{\rho}_3\hat{\sigma}_0;\hat{\cal G}\right]\\&
		-\left[\hat{\Sigma}\hat{\tau}_3\hat{\rho}_3\hat{\sigma}_0;\hat{\cal G}\right]=0, 
	\end{split}
	\en
	where $[\hat{X};\hat{Y}]$ stands for the commutator of matrices and ${\mathbf v}_F=v_F{\mathbf n}$. In equations above, we have omitted writing the dependence on $\bR$ and $\omega_n$ in relevant functions for brevity. Importantly, since the quasiclassical equations are linear in $\hat{\cal G}$ one needs to specify the constraint condition for this function to avoid an ambiguity. Simple calculation shows that the quasiclassical function must satisfy the nonlinear normalization condition
	\beg\label{Norm}
	\hat{\cal G}^2=\hat{1}.
	\en
	In addition, given the problem at hand, the Eilenberger equation above needs to be supplemented with the boundary conditions.

	%###############################################################################################################
	\subsection{Boundary conditions}
	%\tcr{In what follows, we will develop the solution of the Eilenberger equation (\ref{Eq2}) for the Josephson contact between two superconductors allowing for non-zero spin-density-wave order parameter, $M\not=0$. We assume that the normal to the interface between the two superconductors is along the $x$-axis, Fig. \ref{Fig-Contact}, so that all functions in (\ref{Eq2}) are functions of $x$-component of the vector $\bR$ only.} \tcg{We never develop solution in this section. This is out of place should be deleted}
	
	To determine the spacial variation of the order parameters $\Delta(x)$ and $M(x)$ through the interface, we need to solve (\ref{Eq2}) on each side of the interface and then match the quasiclassical functions at the interface ($x=0$) with the use of the boundary conditions \cite{Zaitsev1984}:
	\beg\label{BCs2}
	{\hat{\cal G}_a(0)\left[R\hat{\cal G}_{s+}^2(0)+\hat{\cal G}_{s-}^2(0)\right]=D\hat{\cal G}_{s-}(0)\hat{\cal G}_{s+}(0)}.
	\en
	Here the dependence of the quasiclassical functions on Matsubara frequency has been suppressed,
	$\hat{\cal G}_{s(a)}(0)=(\hat{\cal G}(v_x,0)\pm\hat{\cal G}(-v_x,0))/2$ and 
	$\hat{\cal G}_{s\pm}(0)=(\hat{\cal G}_{s}(+0)\pm\hat{\cal G}_{s}(-0))/2$. New parameter $D$ is the transparency coefficient for the interface, while $R=1-D$. 
	Note that the boundary conditions are non-linear. This happens because the interference between the quasiparticle paths on both sides of an interfaces has been completely ignored. The effects of the interference between the trajectories go beyond the scope of this work and will be reported elsewhere \cite{PathInterfere}.

	%###############################################################################################################
	\section{Analysis of the quasiclassical equations}
	In this Section, we first discuss the approach to solving the Eilenberger equations and then show the results of our numerical solution for the
	spatial variation of $\Delta(x)$ and $M(x)$ across the junction. 
	\subsection{Ansatz for the quasiclassical functions}
	The task of solving matrix Eilenberger equation (\ref{Eq2}) presents a major challenge. In addition to being supplemented by the nonlinear boundary conditions (\ref{BCs2}) for a given $\Delta(x)$ and $M(x)$, these functions  must be determined self-consistently via relations \cite{Josephsonic2018}:
	\beg\label{self-consistent}
	\begin{split}
		\frac{\Delta(x)}{\nu_F\lambda_{\rm{sc}}}&=2\pi T\sum^{\Omega_\Lambda}_{\omega_n}\mathrm{Tr}\left[\langle\hat{\cal G}\rangle(\hat{1}+\hat{\tau}_3)\hat{\rho}^{+}\hat{\sigma}^{+}\right],\\
		\frac{M(x)}{\nu_F\lambda_{\rm{sdw}}}&=2\pi T\sum^{\Omega_\Lambda}_{\omega_n}\mathrm{Tr}\left[\langle\hat{\cal G}\rangle\hat{\tau}^+(\hat{1}+\hat{\rho}_3)\hat{\sigma}_3\right],
	\end{split}
\end{equation} 
where $\langle\hat{\cal G}\rangle=\langle\hat{\cal G}(x,\omega_n,v_x)\rangle$ and averaging is performed over the directions of the quasiparticle trajectories. Here, $\Omega_\Lambda$ is the energy scale of an ultraviolet cutoff, while $\lambda_{\textrm{sc},\textrm{sdw}}$ are the corresponding coupling constants and we employed the standard notation $a^+=a_x+ia_y$. 

Clearly, to make further progress we need to specify the matrix structure of the function $\hat{\cal G}$ that will respect the nonlinear normalization constraint. We now present the ansatz for the function $\hat{\cal G}$, that follows the constraint: 
\beg\label{G}
\hat{\cal G}(\zeta)=g_z(\zeta){\hat{\tau}_3\hat{\rho}_3\hat{\sigma}_0}+\hat{\cal G}_{\textrm{sc}}(\zeta)+\hat{\cal G}_{\textrm{sdw}}(\zeta)+\hat{\cal G}_{\textrm{mix}}(\zeta).
\en
where $\zeta=(x,\omega_n, v_{F}n_x)$. The first term in (\ref{G}) is the quasiclassical function for the normal component, so that in the absence of an interface and when $\Delta=M=0$ it obtains $g_z(i\omega_n)=\textrm{sign}(\omega_n)$. The second term accounts for the superconducting correlations:
\beg\label{Gsc}
\hat{\cal G}_{\textrm{sc}}(\zeta)=f_{z}(\zeta){\hat{\tau}_0\hat{\rho}_1\hat{\sigma}_2}+if_x(\zeta){\hat{\tau}_3\hat{\rho}_2\hat{\sigma}_{2}}.
\en
Here the anomalous $f_z$ component must be constant in the bulk, while $f_x$ is only nonzero in a close proximity to an interface and is an odd function of unit vector $n_x$. Similarly 
\beg\label{Gsdw}
\hat{\cal G}_{\textrm{sdw}}(\zeta)=s_z(\zeta){\hat{\tau}_2\hat{\rho}_0\hat{\sigma}_3}+
is_x(\zeta){\hat{\tau}_1\hat{\rho}_3\hat{\sigma}_3}.
\en
Finally, the last term in (\ref{G}), as will be shown below,  appears only when both $\Delta\not=0$ and $M\not= 0$:
\beg\label{Gmix}
\hat{\cal G}_{\textrm{mix}}(\zeta)=-ig_x(\zeta){\hat{\tau}_2\hat{\rho}_1\hat{\sigma}_1}
\en
and it vanishes in the bulk on the both sides of an interface. 

After substituting these expressions into (\ref{Eq2}) and equating the terms proportional to the same combination of the direct matrix products
$\hat{\tau}_i\hat{\rho}_j\hat{\sigma}_k$ we find the following equations:
\beg\label{Equations12}
\begin{split}
	&\Pi_zs_x-\Phi_z f_x+\Pi_x s_z+\Phi_xf_z-\frac{v_x}{2}\frac{\partial g_z}{\partial x}=0, \\
	&\Sigma_zf_z-\Phi_zg_z+\Pi_xg_x-\Theta_x s_x-\frac{v_x}{2}\frac{\partial f_x}{\partial x}=0, \\
	&\Sigma_zf_x-\Phi_x g_z+\Pi_zg_x+ \Theta_x s_z-\frac{v_x}{2}\frac{\partial f_z}{\partial x}=0, \\
	&\Sigma_zs_z-\Pi_zg_z-\Theta_x f_x-
	\Phi_x g_x+\frac{v_x}{2}\frac{\partial s_x}{\partial x}=0, \\
	&\Sigma_zs_x+\Phi_zg_x+\Pi_x g_z+\Theta_x f_z+\frac{v_x}{2}\frac{\partial s_z}{\partial x}=0, \\
	&\Pi_zf_z-\Phi_zs_z-\Pi_xf_x-\Phi_xs_x-\frac{v_x}{2}\frac{\partial g_x}{\partial x}=0.
\end{split}
\en
To keep concise notations we have introduced an additional self-energy functions
\beg\label{SelfEns}\nonumber
\begin{split}
	&\Phi_x=\Gamma_t\langle f_x\rangle, ~ \Pi_x=\Gamma_m\langle s_x\rangle, ~\Pi_z=M(x)-\Gamma_t\langle s_z\rangle, \\
	&\Theta_x=\Gamma_t\langle g_x\rangle,
	~\Sigma_z=\omega_n+\Gamma_t\langle g_z\rangle, ~ \Phi_z=\Delta(x)+\Gamma_m\langle f_z\rangle, 
\end{split}
\en
with $\Gamma_m=\Gamma_0-\Gamma_\pi$ and $\Gamma_t=\Gamma_0+\Gamma_\pi$, and with implicit averaging that is performed over all possible values of unit vector $n_x$.

%###############################################################################################################
\subsection{Quasiclassical function components in the bulk}
Eilenberger equation acquires the simplest form in the bulk when the gradient term can be discarded. For definiteness we consider the bulk of a superconductor at $x>0$. 
According to our discussion above, only three functions $g_z$, $f_z$ and $s_z$ are non-zero. Simple calculation yields
\beg\label{MFEqs}
\begin{split}
	\left(\omega_n+2\Gamma_\pi g_{z}^b\right)f_{z}^b&=\Delta g_{z}^b ,
	~\left(\omega_n+2\Gamma_tg_{z}^b\right)s_z^b=M g_{z}^b, \\
	\left(M-2\Gamma_0 s_z^b\right)f_z^b&=\Delta s_z^b,
\end{split}
\en
where superscript $b$ in all the functions implies value of that function taken in the bulk of a sample, namely $g_z^b=g_z(x\to\infty)$ etc. Furthermore, it is easy to show that the third equation is redundant. However, as we will see below, in the vicinity to the interface an analogue of this equation will determine the spatial variation of the function $g_x$, Eq. (\ref{Gmix}). 

Numerical solution of the first two equations together with the self-consistency equations (\ref{self-consistent}), produces the well known phase diagram of SC-SDW coexistence shown in Fig. \ref{Fig-PhaseDiagram} for a certain choice of parameters (compare that to Refs. \cite{VC-PRB11,London2015,Josephsonic2018}). This model reveals the dome-like structure of superconductivity overlapping with SDW state. Bending of the superconducting dome in the nonmagnetic phase occurs due to finite $\Gamma_\pi$ that serves as an effective pair-breaking factor for $s^{\pm}$ superconductivity.  Suppression of magnetic order already occurs at the level of intra-band scattering that is governed by $\Gamma_0$. The width of the coexistence region $\mathrm{max}[T_N(\Delta)-T_c(M)]$ can be controlled by the ratio between the scattering rates $\Gamma_\pi/\Gamma_0$ and can change substantially, however within this model it always remains rather narrow.   
 
 %###############################################################################################################
	\begin{figure}[h]
		\includegraphics[width=8cm]{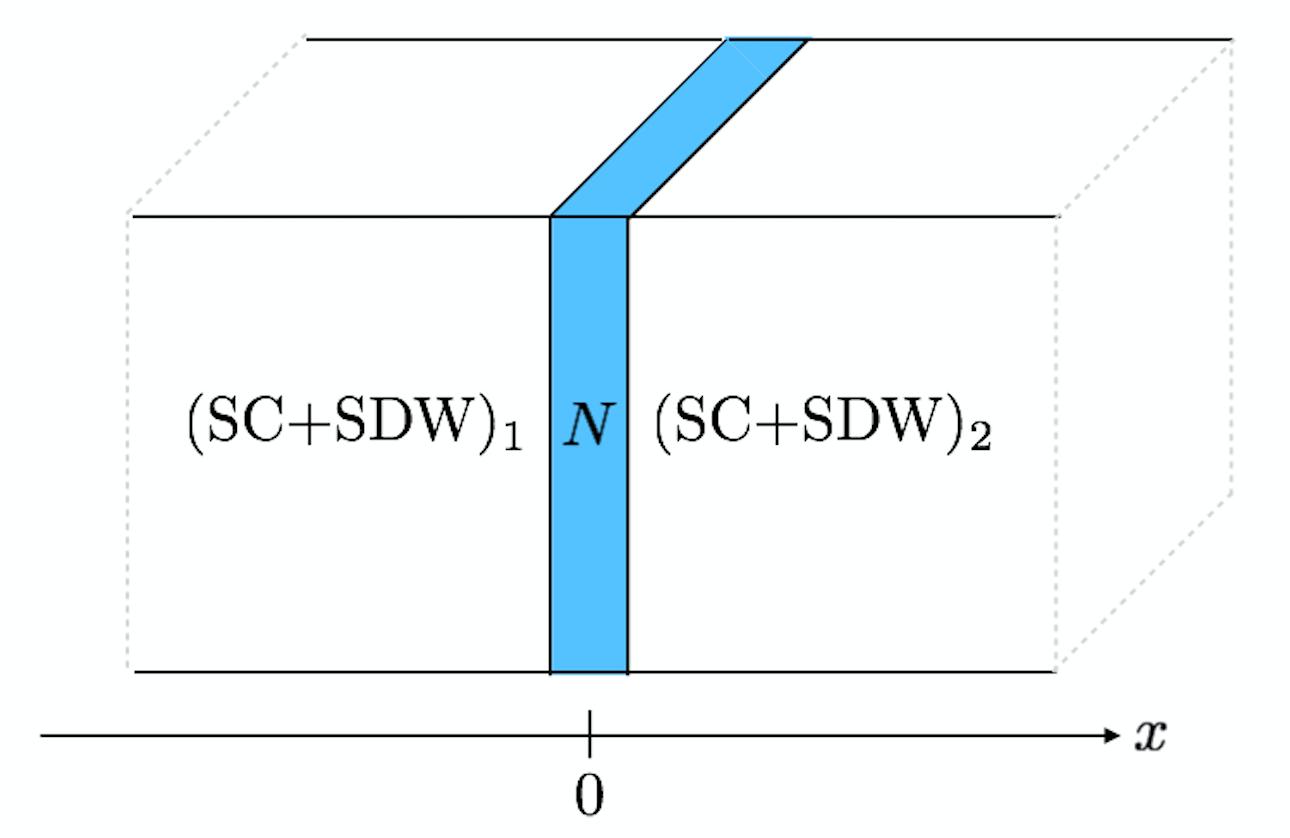}
		\caption{Schematic representation of the Josephson contact between two superconductors separated by a metallic barrier. It is assumed that in both superconductors magnetic (i.e. spin-density-wave) order parameter can be nonzero.}
		\label{Fig-Contact}	
	\end{figure}
%#############################################################################################

\subsection{Normalization condition}
Simple algebraic manipulations with equations (\ref{Equations12}) show that components of $\hat{\cal G}$ satisfy
\beg\label{norm2}
g_z^2+f_z^2+s_z^2-f_x^2-s_x^2-g_x^2=\textrm{const.}
\en
for any value of $x$. By sending $x\to\pm\infty$, it immediately follows that the constant must be equal to one. 
On the other hand we can use Eq. (\ref{Norm}) directly with (\ref{G}) to find
\beg\label{OhOh}
\hat{\cal G}^2=\hat{1}-2i\left(g_zg_x-f_zs_x-f_xs_z\right){\hat{\tau}_1\hat{\rho}_2\hat{\sigma}_1},
\en
where we already took into account (\ref{norm2}). Again, as it can be checked by the direct calculation the second term here is actually a constant
\beg\label{Constraint2}
{g_zg_x-f_zs_x-f_xs_z=\textrm{const}}.
\en
Constant appearing in this equation must be zero due to the vanishing of functions $g_x$, $s_x$ and $f_x$ in the bulk. Thus, we have derived the matrix form of the quasiclassical function and have  demonstrated that normalization condition for the function $\hat{\cal G}$ holds.

%###############################################################################################################
\begin{figure}[h]
	\includegraphics[width=8cm]{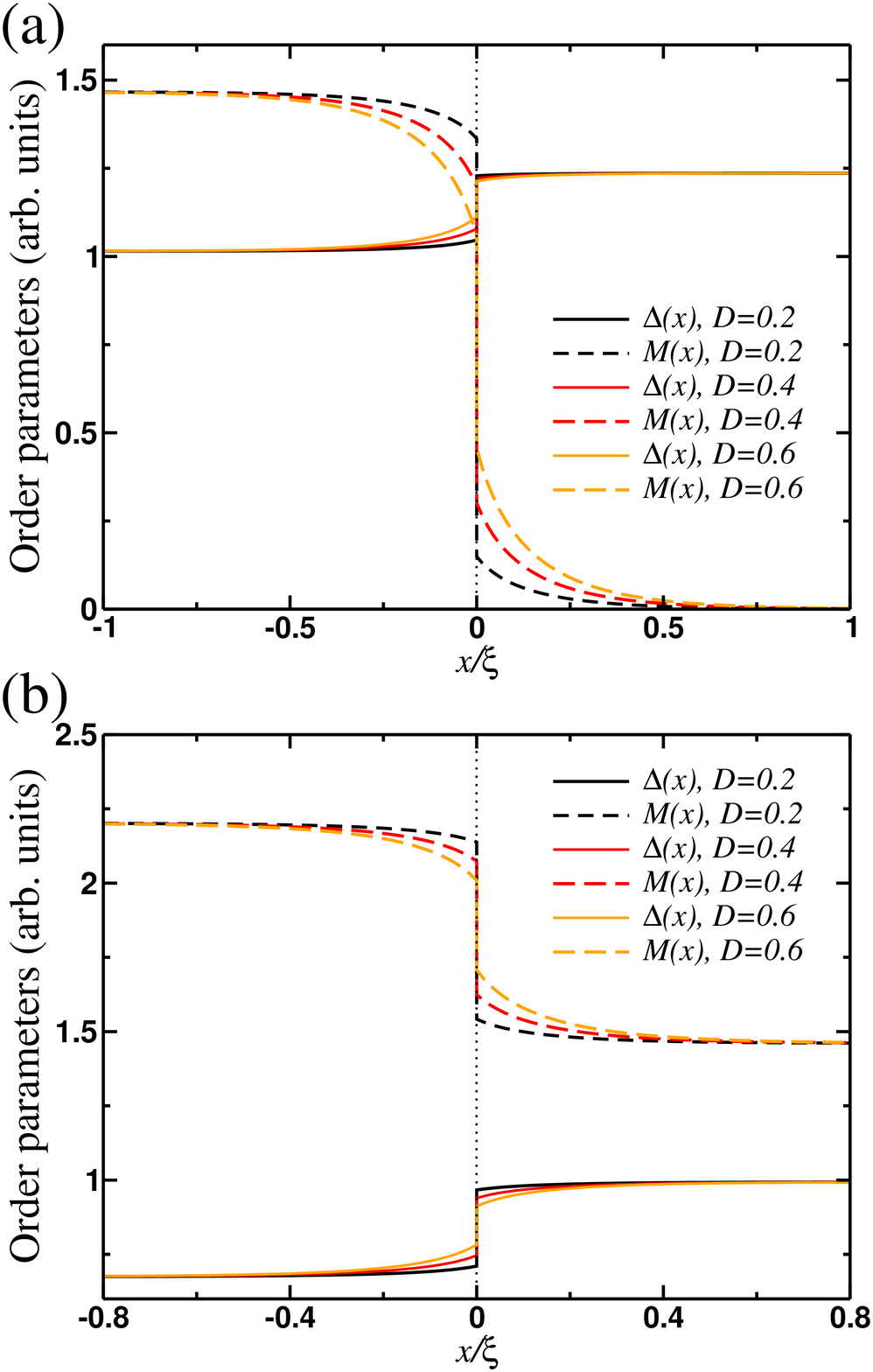}
	\caption{(color online) Results for the spatial dependence of the order parameters obtained from the numerical solution of the Eilenberger equations for various values of the transmission coefficient $D=1-R$. The distance away from the contact is normalized to the coherence length. }
	\label{Fig-DltxMx}	
\end{figure}
%###############################################################################################################

\section{Applications and results}
In this Section, we present the results of our analysis of the quasiclassical equations and use these results to compute the observables: local density of states and critical current. 
\subsection{Order parameters and local density of states}

The numerical solution of the quasiclassical equations (\ref{Equations12}) for the geometry of a junction illustrated in Fig. \ref{Fig-Contact}, is plotted in Fig. \ref{Fig-DltxMx}. These plots are one of the main results of this paper concerning the nature of the proximity effect in a complex superconducting phase. One important observation that we can make in regards to the spacial changes of the superconducting order parameter is that it varies substantially only in an immediate vicinity of the boundary between the two superconductors. On the contrary, the spin-density-wave order parameter changes on a somewhat larger length scale. Therefore, our results formally justify the often used approximation of constant order parameter on both sides of an interface.

It is important to point out here that precisely this aspect of the problem leads to practically universal predictions for the current-phase relations of S$_{\mathrm{SDW}}$-N-S$_{\mathrm{SDW}}$ Josephson junctions. It should be noted, however, that disorder model considered here is not the only one that gives coexistence scenario. In the band models \cite{VVC,FS}, coexistence region can be significantly broader in parameter space so that proximity problem in principle may also have qualitatively different behavior, in particular displaying longer coherence lengths.       
%###############################################################################################################
\begin{figure}
	\includegraphics[width=7cm]{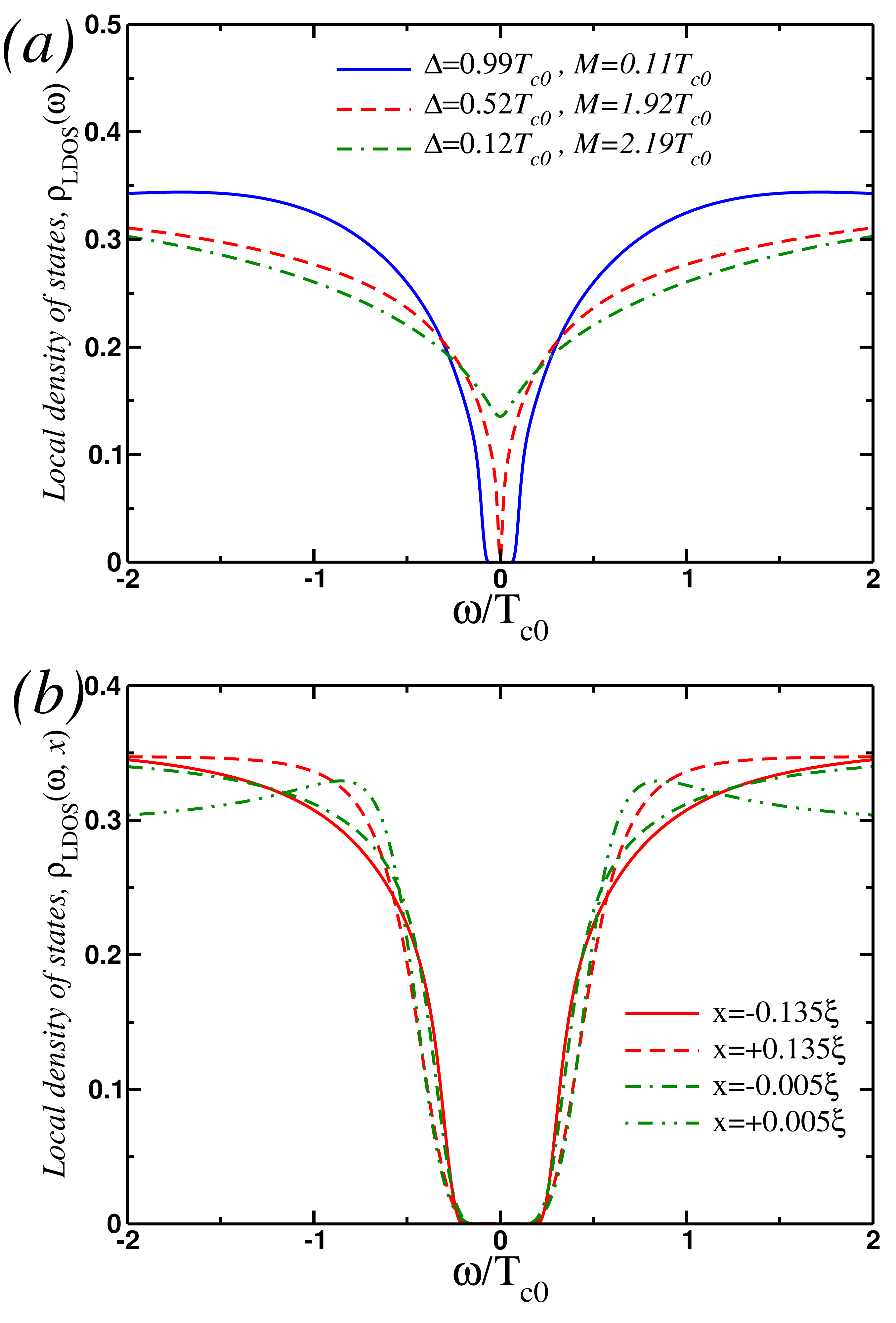}
	\caption{Panel (a): local density of states in the bulk plotted for different choice in values of disorder scattering rates. Panel (b): local density of states plotted
		for various values of the distance (units of the average coherence length of two superconductors) from an interface for two superconductors with the following values of the order parameters in the bulk: $\Delta_1=1.01T_{\textrm{c0}}$, $M_1=1.48T_{\textrm{c0}}$, $\Delta_2=1.24T_{\textrm{c0}}$, $M_2=10^{-4}T_{\textrm{c0}}$. The reflection coefficient of an interface is $R=0.6$. We have set the temperature to $T=0.01T_{c0}$.}
	\label{Fig-LDOS}	
\end{figure}
%###############################################################################################################

\begin{figure}
	\includegraphics[width=7cm]{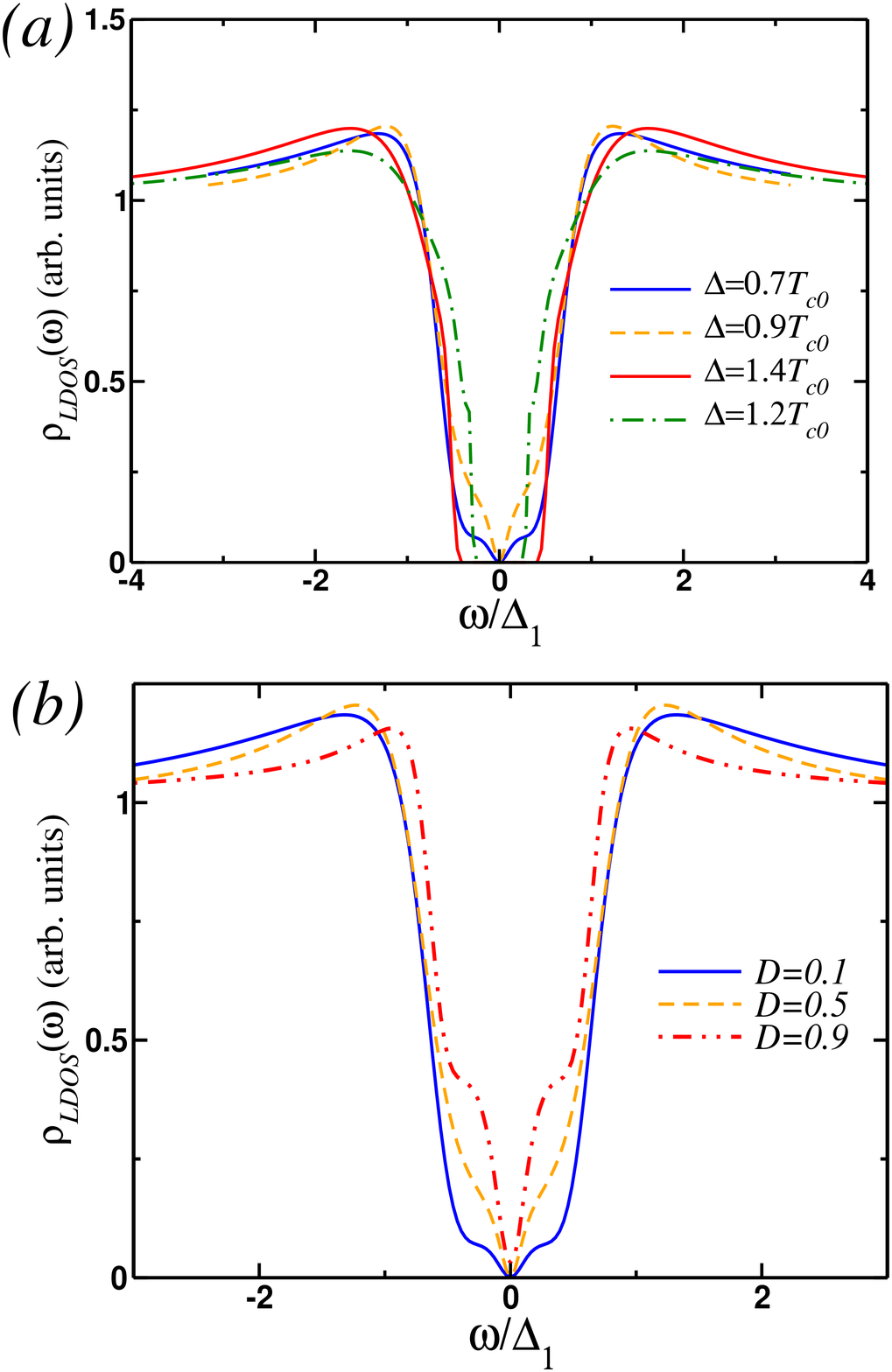}
	\caption{Local density of states in the bulk (top panel) and at the interface (bottom panel) for various values of the disorder scattering rates and interface deflection coefficient for the case of zero magnetization at $T=0.1T_{c0}$. $T_{c0}$ is the superconducting critical temperature in a clean system. The energy is given in the units of $\Delta_1=1.05T_{c0}$.}
	\label{Fig-LDOSM0}	
\end{figure}

With the solution of the Eilenberger equations, we can easily determine the local density of states at the interface ($x=+0$), using the well known expression \cite{Kiesel1987,Bruder1990}
\beg\label{rhoLDOS}
\nonumber
\begin{split}
	\rho_{\textrm{LDOS}}(\omega,x)=\left\langle\textrm{Tr}\left\{\textrm{Re}\left[\hat{\tau}_3\hat{\rho}_3\hat{\sigma}_0\hat{\cal G}(\omega+i0,x,n_x)\right]\right\}\right\rangle
\end{split}
\en
and the averaging is performed over all directions of unit vector ${\vec n}$. For a fixed position from an interface, order parameter $\Delta(x=+0)$ and components of $\hat{\cal G}(i\omega_n,x=+0,n_x)$ are known from the numerical solution. Upon averaging over $n_x$, we obtain $\langle\hat{\cal G}(i\omega_n,x,n_x)\rangle$ and by performing an analytic continuation to real frequencies, $i\omega_n\to\omega+i0$, we then, are able to compute the local density of states by employing the P\'{a}de approximation. In Fig. \ref{Fig-LDOS}, we 
present the results of these calculations for $\rho_{\textrm{LDOS}}(\omega)$ in the bulk superconductor (top panel) and at the interface (bottom panel). Our results for the $\rho_{\textrm{LDOS}}(\omega)$ show several features which appear as a result of non-zero disorder and finite temperatures. In this regard the comparison of the LDOS at the interface with the one in the bulk affords a fairly easy interpretation of our results. In particular, the narrowing of the region near $\omega=0$ describing fairly sharp increase of $\rho_{\textrm{LDOS}}(\omega)$ can be associated with a suppression of the larger pairing amplitude by the presense of the interface and disorder-induced scattering. The similar conclusions can be also drawn from our results for the LDOS for the case of the contact between two superconductors with $M=0$, Fig. \ref{Fig-LDOSM0}.

\subsection{Josephson effects}
In analogy with the local density of states, Josephson current through the junctions also admits representation in terms of the Eilenberger function:
\begin{align}\label{JCPR}
&J=e\nu_Fv_FT\sum\limits_{\omega_n}\int\limits_{0}^{\pi/2}\frac{d\phi}{2\pi}\sin\phi \nonumber\\ 
&\textrm{Tr}\left\{\mathrm{Im}\left[\hat{\tau}_3\hat{\rho}_3\hat{\sigma}_0\hat{\cal G}_a(i\omega_n,0,v_F\sin\phi)\right]\right\}.
\end{align}
In order to compute $\hat{\cal G}_a(0)$, one generally speaking needs to consider equations \eqref{Eq2} with the complex $\Delta(x)$ on each side of the interface. However, in the case when there is no magnetic field the problem can be significantly simplified by using the unitary transformation, which eliminates the phase from the order parameter. For example, assume that the order parameter on the left side of the interface is 
$\Delta_1(x)=|\Delta_1(x)|e^{i\chi(x)}$. Then, we introduce a unitary transformation
\beg\label{Unitary}
\hat{\cal S}(\chi)=\cos(\chi/2)\hat{\tau}_3\hat{\rho}_0\hat{\sigma}_{0}-i\sin(\chi/2)\hat{\tau}_3\hat{\rho}_3\hat{\sigma}_{0}.
\en
It is easy to verify that $\hat{\cal S}\Delta_1\hat{\cal S}\dg=|\Delta_1|$. If one now implements this unitary transformation for the quasiclassical 
functions, it follows that the Eilenberger equation acquires essentially the same form as the one with purely real order parameter, except for the extra term proportional to $\partial\chi/\partial x$. This term, however, can be ignored for one does expect the phase to vary substantially across the junction. Furthermore, by performing the inverse unitary transformation, one can determine function $\hat{\cal G}_a(i\omega_n,0,v_F\sin\phi)$. 

\begin{figure}[b!]
\includegraphics[width=8cm]{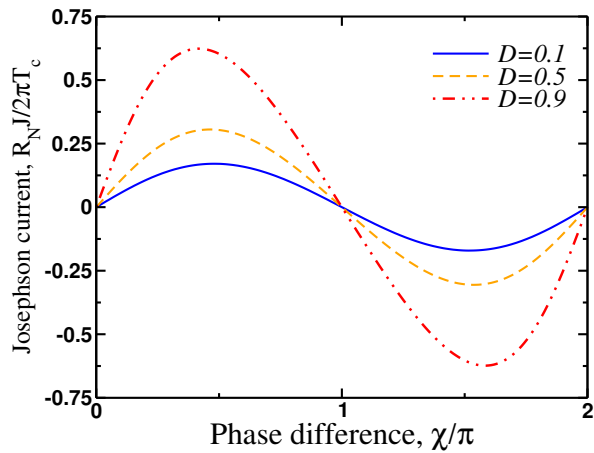}
\caption{Josephson current through the junction as a function of the phase difference between the superconducting order parameters on the both sides of the interface for different values of the transparency coefficient $D$. The order parameter $\Delta$ was computed self-consistently and $R_N$ is the resistance of the junction in the normal state.}
\label{Fig:JCPR}	
\end{figure}

After somewhat lengthy calculation, we found
\begin{widetext}
\begin{equation}\label{JCPR-Trace}
{\textrm{Tr}\left[\hat{\tau}_3\hat{\rho}_3\hat{\sigma}_0\hat{\cal G}_a(0)\right]=  
\frac{2i{D}\left[i\left(f_{1b}^rf_{2b}^l-f_{2b}^rf_{1b}^l\right)\cos\chi+\left(f_{1b}^rf_{1b}^l-f_{2b}^rf_{2b}^l\right)\sin\chi\right]}{2-D\left[1-g_b^rg_b^l+\left(f_{1b}^rf_{1b}^l-f_{2b}^rf_{2b}^l\right)\cos\chi
-i\left(f_{1b}^rf_{2b}^l-f_{2b}^rf_{1b}^l\right)\sin\chi\right]}}.
\end{equation}
\end{widetext}
Here $\chi$ is the global phase difference between the order parameters on the both sides of the interface and the superscripts $l$ and $r$ mean that the functions should be evaluated either on the left ($x=-\delta$) or the right ($x=+\delta$) sides of the interface. Our results for the Josephson current-phase relation are shown in Fig. \ref{Fig:JCPR}. Quite naturally, we find that for small $D$, the Josephson current will be proportional to $\sin\chi$. 
We note that the quasiclassical functions which account for the magnetic order do not explicitly enter into the expression for the Josephson current. 

Motivated by ideas and practical implementation of Josephson Scanning Tunneling Spectroscopy (JSTS) as a diagnostic of unconventional superconductivity \cite{Balatsky2001,NaamanPRL2001,SuderowSST2014,Hamidian2016,YazdaniPRB2016,GrahamPRL2019}, we briefly consider this effect in our model. To determine the dependence of the critic current on external voltage $V$ we use the usual expression \cite{KulikYanson,Balatsky2001}:
\beg\label{KulikJc}
\begin{split}
	&{I}_c(V)=-\frac{1}{4\pi^3eR_N}\iint_{-\infty}^{+\infty}d\omega_1 d\omega_2
	\\&\times\left[\frac{{\cal F}_1\dg(\omega_1){\cal F}_2(\omega_2)}{\omega_1+\omega_2+eV-i\delta}+\frac{{\cal F}_1^*(\omega_1){\cal F}_2^{\dagger*}(\omega_2)}{\omega_1+\omega_2-eV-i\delta}\right].
\end{split}
\en
Here ${\cal F}_{1(2)}(\omega)$ are the quasiclassical anomalous Green's functions for the first (second) superconductor and $R_N$ is the contact resistance. In the JSTS setup, one usually uses SC tip with known properties as a reference point and another SC as a study system. For this reason, we choose ${\cal F}_1(\omega)$ in the form which describes a clean BCS superconductor with the pairing amplitude $\Delta_{\textrm{BCS}}$: 
\beg\label{F1}
{\cal F}_1(\omega)={\cal F}_1\dg(\omega)=\left\{
\begin{matrix}
	\frac{\pi \Delta_{\textrm{BCS}}}{\sqrt{\Delta_{\textrm{BCS}}^2-\omega^2}}, \quad |\omega|<\Delta_{\textrm{BCS}}, \\
	\frac{i\pi \Delta_{\textrm{BCS}}}{\sqrt{\omega^2-\Delta_{\textrm{BCS}}^2}}, \quad |\omega|>\Delta_{\textrm{BCS}}.
\end{matrix}
\right.
\en

%###############################################################################################################
%###############################################################################################################

As for the function ${\cal F}_2(\omega_2)$, it can be directly obtained from $f_z^b(i{\omega_n})$ by performing an analytic continuation. For simplicity we consider $f_z^b(i{\omega_n})$
calculated away from the contact that creates spatial inhomogeneity, but obviously the calculation can be done for any spatial location of the tunneling tip with respect to the junction.  

%###############################################################################################################
\begin{figure}[h]
	\includegraphics[width=8cm]{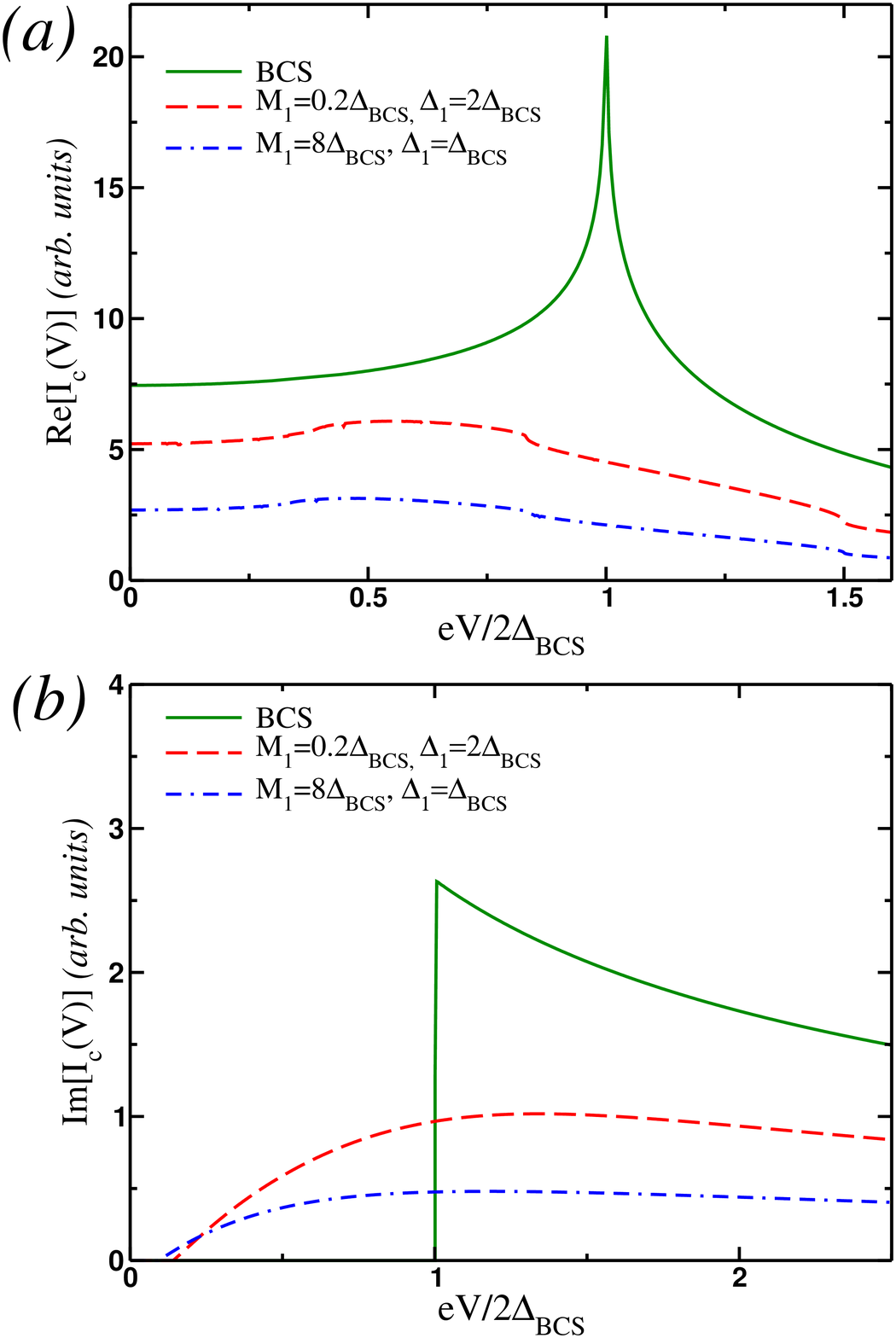}
	\caption{Plot of the real and imaginary parts of the Josephson critical current through the junction as a function of external voltage. Panel (a): the real part of the critical current as a function of external voltage in the units of the BCS gap. The solid line correspond a junction between two BCS superconductors, while the remaining two lines are for the junction when one of the superconductors is in co-existence with the SDW order. Panel (b): the imaginary part of the critical current.}
	\label{Fig-Ic}	
\end{figure}
%###############################################################################################################

The results of our numerical calculations based on Eq. \eqref{KulikJc}, are presented on Fig. \ref{Fig-Ic}. To make a contrast with the textbook example, we also plotted the critical current for a contact between two BCS superconductors. It is worth reminding the reader that in the case of two identical BCS superconductors, the real part of the critical current has a logarithmic divergence at the threshold voltage $eV=2\Delta_{\textrm{BCS}}$, while the imaginary part has a square-root singularity at the same value of external voltage. Now, if one of the BCS superconductors is replaced with an unconventional disordered superconductor, the presence of disorder and nonzero magnetization produces the smearing of the above mentioned singularities and lead to the substantial broadening of the dependence $I_c(V)$. 

%###############################################################################################################
\section{Overview of universal Josephson current-phase relations}

The physics of the dc-Josephson effect is ultimately related to the sub-gap states that carry the supercurrent. These states form as a result of Andreev reflections that electrons undergo when impinging on superconducting interfaces.  Location of these states inside the superconducting gap depends on the superconducting phase difference across the junction. Kulik solved the first microscopic model of superconductor-normal-superconductor (SNS) model and derived the spectrum of Andreev states in various limits (see e.g. Refs. \cite{Golubov-RMP,KulikYanson}). In a way, this work marked the beginning of intensive studies of various kinds of Josephson junctions that spanned over multiple decades. The interest to this problem has been continuously sustained to the present day not only due to the fundamental physics involved and applications of this effect, but also emergence of the new classes of unconventional superconductors.

The most elegant way to derive the spectrum of Andreev bound states is by using scattering matrix approach.
Beenakker derived the general determinant formula \cite{Beenakker-3UJE}, which has very transparent and intuitive meaning. In the limit of the short junction, when length of the link separating superconducting leads is small compared to the coherence length ($L\ll\xi$), this formula simplifies to a famous expression for a pair of Andreev levels per-channel of the junction 
\begin{equation}\label{ABS}
	E_n(\chi)=\pm\Delta\sqrt{1-D_n\sin^2(\chi/2)}.
\end{equation}
In this theory, the junction is modeled as a multi-mode conductor where each conduction channel labeled by an index $n$ has certain transmission coefficient $D_n$. 
The Josephson current $J(\chi)$ carried by these states is given by
\begin{equation}\label{J-ABS}
	J(\chi)=\frac{e\Delta^2}{2\hbar}\sin(\chi)\sum_n\frac{D_n}{E_n}\tanh\left(\frac{E_n}{2T}\right). 
\end{equation}
With this formula at hand, one can recover multiple special cases. Indeed, at temperatures close to the critical, $T\lesssim T_c$, the superconducting gap is small, $\Delta\ll T$, so that one can expand the thermal factor of hyperbolic tangent at small argument, which gives as a result  
\begin{equation}
	J(\chi)=\frac{\pi\Delta^2}{4eR_NT}\sin(\chi),
\end{equation}      
where we introduced the total normal state resistance of the junction by means of the Landauer formula $R^{-1}_N=(2e^2/h)\sum_nD_n$. This sinusoidal current-phase relationship for a superconductor-constriction-superconductor (ScS) junction was originally derived by Aslamazov and Larkin from the Ginzburg-Landau theory \cite{AslamazovLarkin1976}. This result is universal in the sense that it applies to any kind of constriction. It is also a generic property that Josephson current is harmonic (sinusoidal) near $T_c$.  Alternatively, one can consider a tunnel barrier, $D_n\ll1$, which corresponds to a class of superconductor-insulator-superconductor (SIS) type junctions. This yields the following expression for the current in the form 
\begin{equation}
	J(\chi)=\frac{\pi\Delta}{2eR_N}\sin(\chi)\tanh(\Delta/2T). 
\end{equation}
that was derived first by Ambegaokar and Baratoff from the tunneling Hamiltonian \cite{Ambegaokar1963}. In the opposite limit of fully transparent channels, $D_n=1$ for $n=1,\ldots,N$, one recovers the model of quantum point contact, namely S-QPC-S junction. In this case, the spectrum of Andreev levels simplifies to $E_n=\Delta\cos(\chi/2)$ for any channel and the corresponding current is      
\begin{equation}
	J(\chi)=\frac{\pi\Delta}{eR_N}\sin(\chi/2)\tanh\left[\frac{\Delta\cos(\chi/2)}{2T}\right].
\end{equation}
This formula was derived first by Kulik and Omelyanchuk from the Eilenberger equations \cite{KulikOm1975}. 

In realistic contacts of actual devices, transmissions are neither fully ballistic nor of tunneling type, rather there is a continuous distribution $\rho(D)$ of transmission eigenvalues. There are several generic contact types that have been discussed in the literature. Their distributions are described by the function of the form
\begin{equation}\label{DTE}
	\rho(D)= N_p\frac{1}{R_NG_Q}\frac{1}{D^{p}\sqrt{1-D}},\quad p=1/2,1,3/2
\end{equation}   
The case with the power exponent $p=1/2$ corresponds to two ballistic connectors with equal conductances in series. The case with $p=1$ corresponds to the Dorokhov function of a diffusive connector \cite{Dorokhov}. The case with $p=3/2$ was considered by Schep and Bauer \cite{Schep} and corresponds to an interface with a high density of randomly distributed scatterers. The normalization factors $N_p$ are chosen in such a way as to ensure the total resistance of the junction to be $R^{-1}_N=G_Q\int^{1}_{0}D\rho(D)dD$, where $G_Q$ is the quantum of conductance. It is perhaps remarkable to see that averaging Josephson current in Eq. \eqref{J-ABS}, which was derived for a given set of transmissions over their distributions with the help of Eq. \eqref{DTE}, $\sum_n \ldots \to\int^{1}_{0}(\ldots)\rho(D)dD$ reproduces known results, which were obtained by means of semiclassical technique. 

Consider $p=1$ first: the normalization is $N_{1}=1/2$, and the ensemble averaged Josephson current is (taken at zero temperature for simplicity)     
\begin{equation}
	J(\chi)=\frac{\pi\Delta}{4eR_N}\int^{1}_{0}\frac{\sin(\chi)dD}{\sqrt{1-D}\sqrt{1-D\sin^2(\chi/2)}}.
\end{equation}
The integral can be found in elementary functions with the final result 
\begin{equation}
	J(\chi)=\frac{\pi\Delta}{eR_N}\cos(\chi/2)\mathrm{arctanh}[\sin(\chi/2)]
\end{equation}
that corresponds to Kulik-Omelyanchuk computation carried out for the disordered SNS junction based on the Usadel equations \cite{KO-Diff}. For the case with $p=3/2$, an analogous averaging yields  
\begin{equation}
	J(\chi)=\frac{\Delta}{2eR_N}\int^{1}_{0}\frac{\sin(\chi)dD}{\sqrt{D(1-D)}\sqrt{1-D\sin^2(\chi/2)}}
\end{equation}
which after the substitution $D=\sin^2(x)$ reduces to the complete elliptic integral of the second kind 
\begin{equation}
	J(\chi)=\frac{\Delta}{eR_N}\sin(\chi)\mathbb{K}\left[\sin(\chi/2)\right]. 
\end{equation}
This current-phase relationship was obtained first by Kupriyanov and Lukichev from Usadel equations in the context of SINIS junction (see 
Ref. \onlinecite{Lukichev1984} for review). Its connection to Eq. \eqref{J-ABS} with subsequent averaging over $\rho(D)$ was pointed out by Brinkman and Golubov \cite{BrinkmanGolubov2000}. 
Lastly, the case with $p=1/2$ corresponds to a chaotic cavity/quantum dot that supports current 
\begin{equation}
	J(\chi)=\frac{\Delta}{eR_N}\sin(\chi)\int^{1}_{0}\frac{\sqrt{D}dD}{\sqrt{1-D}\sqrt{1-D\sin^2(\chi/2)}}
\end{equation}
as was studied by Brouwer and Beenakker \cite{BrowerBee1995}. The corresponding current-phase relationship can be written as a combination of elliptic functions of the first and second kind   
\begin{equation}
	J(\chi)=\frac{4\Delta}{eR_N}\cot(\chi/2)\left[\mathbb{K}[\sin(\chi/2)]-\mathbb{E}[\sin(\chi/2)]\right]. 
\end{equation}
All these examples give different functional form of the Josephson current, yet all of the them support parametrically the same critical current, $J_c\simeq \Delta/eR_N$, which is governed by the total conductance of the junction in the normal state and size of the gap in the leads. In that regard these results are universal. In the extended junctions, when the size of the weak link is large as compared to the coherence length $L\gg\xi$, the situation is different as critical current will decay as a power law or even exponentially with $L$ depending on temperature.  The decay of the current is related to the large dwell time needed for quasiparticles to travel across the junction. Additional features may appear due to the complexities of the proximity effect related to induced spectral gaps, including secondary gaps near $\Delta$, that ultimately modify current amplitude and its dependence on the phase \cite{WhislerPRB2018}.  

Interestingly, the family of such (almost universal) results for mesoscopic systems can be extended to include more complex proximity junctions of correlated electrons, such as S$_{\mathrm{SDW}}$NS$_{\mathrm{SDW}}$ or S$_{\mathrm{SDW}}$INIS$_{\mathrm{SDW}}$. Indeed, thanks to the exact numerical results we have for the spatial profile of the order parameters, one can take the step function model to the leading approximation. Assuming the symmetric case one then finds for the trace in Eq. \eqref{JCPR-Trace} a rather simple analytical expression   
\begin{align}
&\textrm{Tr}\left[\hat{\tau}_3\hat{\rho}_3\hat{\sigma}_0\hat{\cal G}_a(0)\right]=\nonumber \\ 
&\frac{8iD\Delta^2\sin(\chi)}{(2-D)(\omega^2_n+M^2+\Delta^2)+D(\omega^2_n+M^2+\Delta^2\cos(\chi))}
\end{align}
For a multi-mode junction without inter-mode scattering, one can directly average this expression over $\rho(D)$ which gives for the Josephson current a compact formula
\begin{equation}
J(\chi)=\frac{2\pi T}{eR_N}\sum_{\omega_n}\frac{\Delta^2\sin(\chi)}{\omega^2_n+M^2+\Delta^2}\int\frac{D\rho(D)dD}{(2-D)+DP(\chi)}
\end{equation}
where we introduced dimensionless function 
\begin{equation}
P(\chi)=\frac{\omega^2_n+M^2+\Delta^2\cos(\chi)}{\omega^2_n+M^2+\Delta^2}.
\end{equation}
At zero temperature, the Matsubara sum can be converted into an integral over the real axis of continuous frequencies $2\pi T\sum_{\omega_n}\to\int d\omega$ and remaining calculations can be carried out for any of the discussed models of transmission distributions \eqref{DTE}. For instance, for the S$_{\mathrm{SDW}}$INIS$_{\mathrm{SDW}}$ junction (model with $p=3/2$) one finds 
\begin{align}
J(\chi)=\frac{c_p\Delta^2}{eR_N}\frac{\sin(\chi)}{\sqrt{M^2+\Delta^2\cos^2(\chi/2)}}
\end{align}
where $c_p$ is the numerical factor of the order of one. We notice that in the part of the phase diagram where SDW competes with SC, $M\gg\Delta$, supercurrent is suppressed $J_c\sim \Delta^2/(eR_NM)$. Other models of contacts can be analyzed in the similar way.   

\section{Discussion and outlook}
By using the quasiclassical theory of superconductivity, we have performed the fully self-consistent treatment of the Josephson junctions between two two-band superconductors in which nodeless order parameter with $s^{\pm}$-symmetry, coexists with an itinerant SDW order. By solving the corresponding quasiclassical equations for the Eilenberger functions, we have found the variation of the superconducting and magnetic order parameters across the interface with arbitrary transparency. Using the results of the numerical solution, we have computed the local density of states $\rho_{\textrm{LDOS}}$, Josephson current-phase relations $J(\chi)$, as well as dependence of the nonequilibrium critical current on external voltage, $I_c(V)$. The features pertaining to the presence of the magnetic order, are clearly pronounced in the local density of states, on distances of the order of the coherence length from the interface provided that pairing amplitude in at least one of the superconductors exceeds the magnetic order parameter. For the critical current, we find that (i) suppression in the parts of the phase diagram where SDW dominates superconductivity, and (ii) disorder leads to smearing of the various sharp features in $I_c(V)$ found for the contact between BCS superconductors. 

Our work can be further extended in multiple directions. It is of practical importance to consider effects of disorder for more realistic Fermi surfaces including ellipticity for example. It is of special interest to consider three-band models that is the minimal setting for the appearance of nematic order that has to be included in the Eilenberger semiclassical scheme. There is also clear motivation to extend this semiclassical theory to real time axis to address dynamical responses of superconductors with competing orders.         

\section{Acknowledgments}
This work was supported by the National Science Foundation Grants No. DMR-1506547 (A.A.K and M.D.), and, in part, by the U.S. Department of Energy, Basic Energy Sciences, grant DE-SC0016481 (M.D.). The work of A.L. was supported by the U.S. Department of Energy, Office of Science, Basic Energy Sciences, under Awards No. DE-SC0020313 and DE-SC0017888. A.L. also acknowledges hospitality of MPI-PKS where part of the work on this project was performed. 

\begin{appendix}
	
	\section{Method of an auxiliary solution}
	In what follows, we will briefly review a theoretical approach, first proposed by Yip \cite{Yip1997}, that allows one to circumvent the issues associated with the nonlinearity of the boundary conditions. This approach makes it possible to write down the expressions for the quasiclassical functions, which match each other on the interface with the finite reflection coefficient (see also Refs. \cite{BC1,BC2,BC3,BC4,BC5}). The only assumption which goes into making this procedure work is that superconductors extended to distances on which the correlations functions and the corresponding order parameters recover their bulk values. 
%###############################################################################################################	
	\begin{figure}[t]
		\includegraphics[width=8cm]{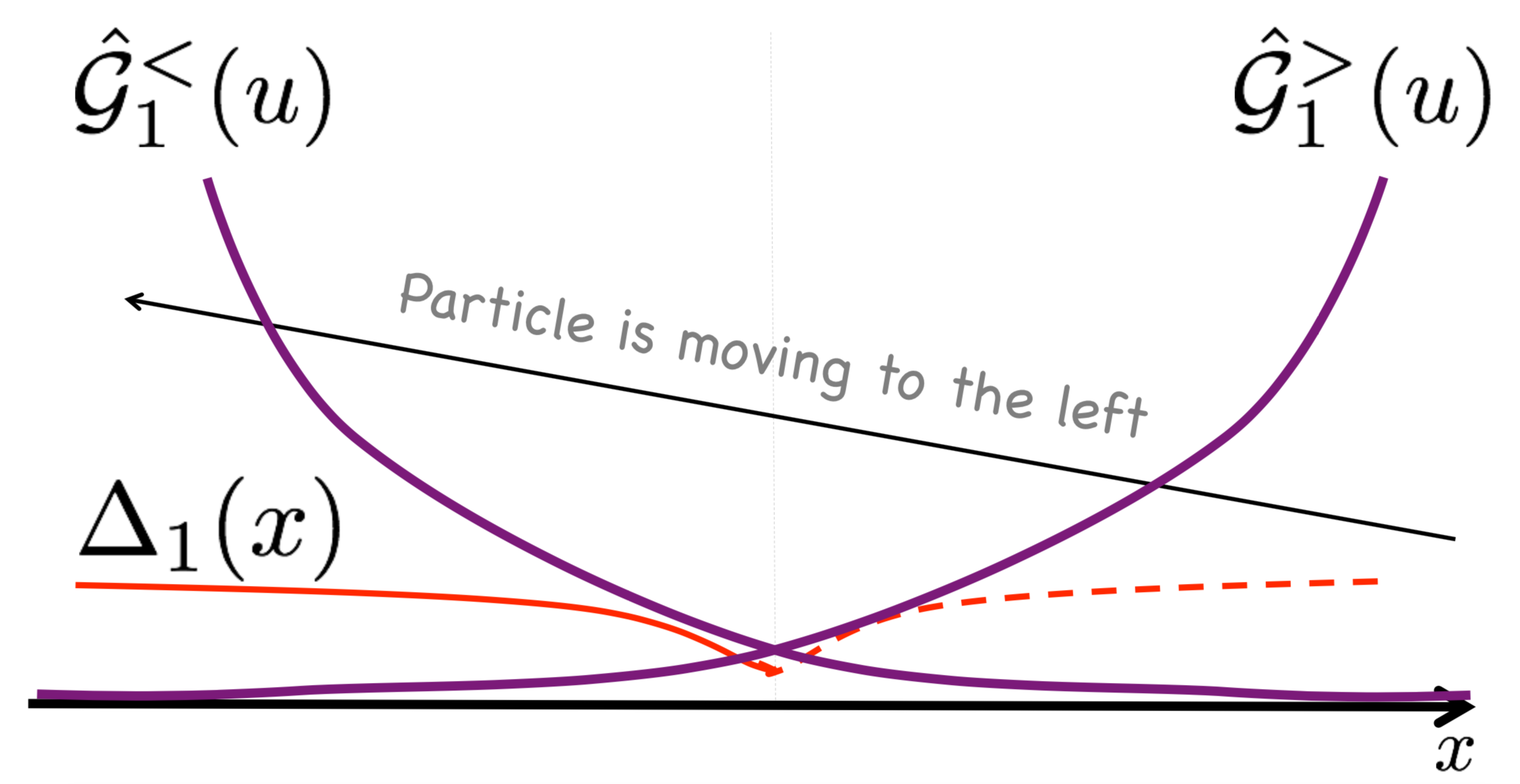}
		\hspace{1cm}
		\includegraphics[width=8cm]{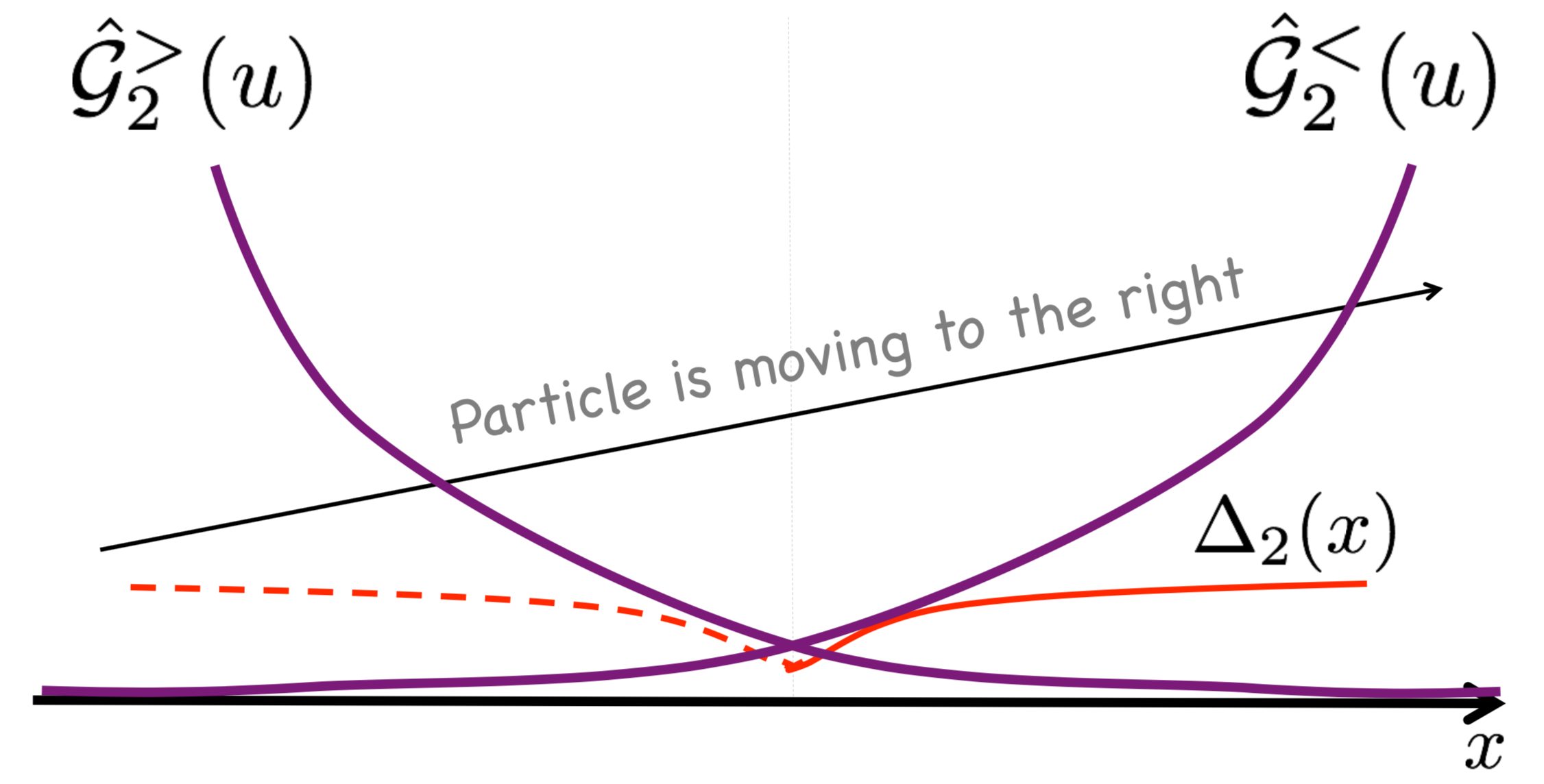}
		\caption{Schematic plot illustrating the procedure of finding two auxiliary solutions of the quasiclassical equations. Auxiliary quasiclassical functions $\hat{\cal G}^{>,<}(u)$ are considered to be the functions of parameter $u=x/v_x\tau_\Delta$, where $\tau_\Delta$ is the relaxation time of the order parameter. Top panel: the trial order parameter is chosen to correspond to the physical order parameter of a superconductor to the left of the interface and the particle's velocity is assumed to be negative, $v_x<0$. Therefore, the diverging solution of the quasiclassical equations at $x\to\infty$ ($u\to-\infty$) is denoted by 
			$\hat{\cal G}_1^{>}(u)$, while the diverging solution at $x\to-\infty$ ($u\to\infty$) is denoted by  $\hat{\cal G}_1^{<}(u)$. Bottom panel: $\hat{\cal G}^{>}(u)$ and $\hat{\cal G}^{<}(u)$ now diverge on opposite sides of the interface and trial form of the order parameter is chosen to match the bulk value of the order parameter on the right hand side of the interface.}
		\label{Fig:Aux12}	
	\end{figure}
%###############################################################################################################	
	The trick we will use, consists of expressing the physical solution in terms
	of the unphysical (i.e. divergent) ones and implementing the resulting relations to simplify the boundary conditions. 
	
	Let us make some general observations. First, it is clear that the Eilenberger equation (\ref{Eq1}) can be formally written as
	\beg\label{EilenFormal}
	v_x{\partial_x \hat{\cal G}}=\left[\hat{\cal L};\hat{\cal G}\right].
	\en
	If $\hat{\cal G}_{sol}$ is a solution of (\ref{EilenFormal}), than ${\hat{\cal G}_{sol}^2}$ is also a solution. Furthermore, due to the normalization 
	condition, it follows that once the boundary conditions are taken into account, our quasiclassical function matrix $\cal{G}$ at the both sides of the interface,
	follows $\hat{\cal G}^2=\hat{\cal G}_0^2$, where ${\cal G}_0$ is a quasiparticle correlator in the bulk, $\hat{\cal G}_0^2=\hat{\cal I}$.  
	
	Instead of solving a problem on both sides of the interface and then trying to match the corresponding solutions, one ignores the interface and solves independently two problems with the parameters matching those at each side of the interface. Solution of each of these two problems, requires a profile of the pairing amplitude as an input, so for the both problems one can use the bulk value of the order parameters at the each side. Within each of these two problems, we need to solve separately for incoming ($n_x>0$) and outgoing ($n_x<0$) trajectories, so generally speaking we are solving four problems in total, Fig. \ref{Fig:Aux12}. 
	
	The logic behind simplifying the boundary condition is as follows. Let us consider two \emph{unphysical} solutions of the Eilenberger equation without an interface, see Fig. \ref{Fig:Aux12}. Specifically, for $v_x>0$ we introduce the physical pairing field in the region $x>0$ and an \emph{auxiliary} field $\hat{\Delta}_{\textrm{aux}}$ in the region $x<0$ and consider
	the solution denoted by  $\hat{\cal G}_{>}\sim e^{-x\lambda/v_x}$ ($\lambda>0$) which diverges (it must diverge since this is an \emph{unphysical} solution) at $x\to-\infty$ and it vanishes as $x\to+\infty$, so that this solution vanishes as $u=x/v_x\to \infty$. 
	Similarly,  $\hat{\cal G}_{<}\sim e^{x\lambda/v_x}$ vanishes at $x\to-\infty$ but it diverges as $x\to+\infty$ with the physical value of the pairing field for $x<0$ and
	auxiliary order parameter for $x>0$. It is easy to show that the product of these two solutions is also a solution of (\ref{EilenFormal}). 
	From the two diverging solutions we can construct the bounded solution:
	\beg\label{Bounded}
	\hat{\cal G}_{\textrm{b}}=a\left(\hat{\cal G}_{<}\hat{\cal G}_{>}-\hat{\cal G}_{>}\hat{\cal G}_{<}\right).
	\en
	Here the normalization constant $a$ is determined by the normalization condition for the physical solution of the Eilenberger equation:
	$a=(\hat{\cal G}_{<}\hat{\cal G}_{>}+\hat{\cal G}_{>}\hat{\cal G}_{<})^{-1}$,
	where we took into account the matrix structure of the quasiclassical functions, i.e. the anticommutator of $\hat{\cal G}_{<}$ and $\hat{\cal G}_{>}$ must be proportional to the unit matrix. Thus, the physical solution (i.e. the one which remains finite in the bulk) in terms of the two unphysical ones reads
	\beg\label{Gb}{
		\hat{\cal G}_{\textrm{b}}=\frac{\hat{\cal G}_{>}\hat{\cal G}_{<}-\hat{\cal G}_{<}\hat{\cal G}_{>}}{\hat{\cal G}_{>}\hat{\cal G}_{<}+\hat{\cal G}_{<}\hat{\cal G}_{>}}}.
	\en
	An important property of these auxiliary matrices is that 
	\beg\label{Zero}
	\hat{\cal G}_>^2=\hat{\cal G}_<^2=0.
	\en
	
	After somewhat lengthy but otherwise straightforward calculation (see below), we obtain the following expressions 
	for the values of the quasiclassical functions at the interface as:
	\beg\label{GaFin}
	\begin{split}
		\hat{\cal G}_a(0)&=\frac{\frac{D}{4}\left[\hat{\cal G}_{\textrm{b}}^r;\hat{\cal G}_{\textrm{b}}^l\right]}{1-\frac{D}{4}\left(\hat{\cal G}_{\textrm{b}}^r-\hat{\cal G}_{\textrm{b}}^l\right)^2}, \\
		\hat{\cal G}_{s}^l(0)&=\frac{\left(1-\frac{D}{2}\right)\hat{\cal G}_{\textrm{b}}^l+\frac{D}{2}\hat{\cal G}_{\textrm{b}}^r}{1-\frac{D}{4}\left(\hat{\cal G}_{\textrm{b}}^r-\hat{\cal G}_{\textrm{b}}^l\right)^2}, \\ 
		\hat{\cal G}_{s}^r(0)&=\frac{\left(1-\frac{D}{2}\right)\hat{\cal G}_{\textrm{b}}^r+\frac{D}{2}\hat{\cal G}_{\textrm{b}}^l}{1-\frac{D}{4}\left(\hat{\cal G}_{\textrm{b}}^r-\hat{\cal G}_{\textrm{b}}^l\right)^2}.
	\end{split}
	\en
	where coordinate dependence on the r.h.s. is suppressed. Remarkably, these relations have the form of the circuit-theory boundary conditions of Andreev refection \cite{Nazarov1994,Nazarov}.
	Thus we were able to express the values of the quasiclassical functions, which determine the physical properties of the junctions, in terms
	of the correlations functions found using the auxiliary solutions. In principle, one can use these values to setup the boundary value problem and solve the Eilenberger equations anew. Indeed, the general solution of the quasiclassical equations with given $\Delta(x)$ can always be written as a linear combination of the bulk solution and the solution of the auxiliary problem, (see Eq. (\ref{Grout}) of Appendix D).
	
	\section{Auxiliary solution in clean junctions}
	To illustrate the power of this method, let us obtain the two auxiliary solutions $\hat{\cal G}^>(u)$ and $\hat{\cal G}^<(u)$ (here $u=x/v_x$ is an auxiliary parameter) of the Eilenberger equations (\ref{Equations12}) in the clean case assuming finite $M$ and then results can be generalized for finite disorder. In equations (\ref{Equations12}) we set $\Gamma_0=\Gamma_\pi=0$ and assume that both $\Delta$ and $M$ are spatially homogeneous. One obtains
	\beg\label{Equations0}
	\begin{split}
		&Ms_x-\Delta f_x=\frac{v_x}{2}\frac{\partial g_z}{\partial x}, 
		~\omega_nf_z-\Delta g_z=\frac{v_x}{2}\frac{\partial f_x}{\partial x}, \\
		&\omega_nf_x+M g_x=\frac{v_x}{2}\frac{\partial f_z}{\partial x}, ~Mg_z-\omega_ns_z=\frac{v_x}{2}\frac{\partial s_x}{\partial x}, \\ 
		&-\omega_ns_x-\Delta g_x=\frac{v_x}{2}\frac{\partial s_z}{\partial x}, 
		~Mf_z-\Delta s_z=\frac{v_x}{2}\frac{\partial g_x}{\partial x}.
	\end{split}
	\en
	Let us now consider two cases of diverging and converging solutions at $x\to\infty$  separately. 
	\paragraph{$\hat{\cal G}^<(x,v_x>0)$} Let us assume $v_x>0$ and focus on the solution of these equations which diverges at $x\to\infty$. We
	look for the solution in the following form 
	\beg\label{Anzats}
	\begin{split}
		f_x^<=a_x^<e^{2\lambda x/v_x}, \quad s_x^<=b_x^<e^{2\lambda x/v_x}, \quad g_x^<=c_x^<e^{2\lambda x/v_x},  \\
		f_z^<=a_z^<e^{2\lambda x/v_x}, \quad s_z^<=b_z^<e^{2\lambda x/v_x}, \quad g_z^<=c_z^<e^{2\lambda x/v_x}.
	\end{split}
	\en
	It follows that the solution can be written as
	\beg\label{cxcz}
	c_x^<=c_1, \quad c_z^<=c.
	\en
	Where $c$ and $c_1$ are two arbitrary constants. The remaining four coefficients are
	\beg\label{axbxazbz}
	\begin{split}
		&a_x^<=-c\frac{\lambda_n\Delta}{M^2+\Delta^2}+c_1\frac{M\omega_n}{M^2+\Delta^2}, \\
		&b_x^<=c\frac{\lambda_n M}{M^2+\Delta^2}b_z+c_1\frac{\Delta \omega_n}{M^2+\Delta^2}, \\
		&a_z^<=-c\frac{\Delta\omega_n}{M^2+\Delta^2}+c_1\frac{M\lambda_n}{M^2+\Delta^2}, \\
		&b_z^<=-c\frac{\omega_n M}{M^2+\Delta^2}-c_1\frac{\Delta \lambda_n}{M^2+\Delta^2}.
	\end{split}
	\en
	Where $\lambda_n=\sqrt{\omega_n^2+\Delta^2+M^2}$. This nontrivial combination for the coefficients must satisfy additional constraints that we will discuss later in this appendix. 
	
    Therefore, for $M\not=0$ two out of six coefficients remain undetermined since the normalization relation \ref{Zero}
	is satisfied identically:
	\beg\label{ZeroNorm}
	a_z^2+b_z^2+c_z^2-a_x^2-b_x^2-c_x^2=0.
	\en

   \paragraph{$\hat{\cal G}^>(x,v_x>0)$} Consider now the case of $\hat{\cal G}^{>}(u\to\-\infty)\to\infty$. Let us again set $v_x>0$. In this case, in Eq. (\ref{Equations0}), we will have to replace $\lambda_n\to -\lambda_n$ to have a divergent solution at $x\to-\infty$, whose coefficients are given by:
	\beg\label{CoeffMore}
	\begin{split}
		&a_x^>=c\frac{\lambda_n\Delta}{M^2+\Delta^2}-c_1\frac{M\omega_n}{M^2+\Delta^2}, \\
		&b_x^>=-c\frac{\lambda_n M}{M^2+\Delta^2}b_z-c_1\frac{\Delta \omega_n}{M^2+\Delta^2}, \\
		&a_z^>=-c\frac{\Delta\omega_n}{M^2+\Delta^2}+c_1\frac{M\lambda_n}{M^2+\Delta^2}, \\
		&b_z^>=-c\frac{\omega_n M}{M^2+\Delta^2}-c_1\frac{\Delta \lambda_n}{M^2+\Delta^2}\\
		&c_x^>=-c_1, \quad c_z=c.
	\end{split}
   \en
	These choices of the coefficients produces the correct value of the quasiclassical propagator in the bulk. 
	
	\section{Bounded solution for clean junctions with spatially homogeneous order parameters}
	The bounded solution is given by Eq. (\ref{Gb}).
	The matrix in the denominator is proportional to the unit matrix 
	\beg\label{Denom}
	\begin{split}
		&\hat{\cal G}^{>}\hat{\cal G}^{<}+\hat{\cal G}^{<}\hat{\cal G}^{>}=-2i{\cal P}{\hat{\tau}_1\hat{\rho}_2\hat{\sigma}_1}+2{\cal Z}{\hat{\tau}_0\hat{\rho}_0\hat{\sigma}_0}.
	\end{split}
	\en
	Where ${\cal P}=g_z^> g_x^<+g_z^<g_x^>-f_z^> s_x^<-f_z^< s_x^>-s_z^> f_x^<-s_z^< f_x^>$. Which goes to zero because $a_x,~b_x$ and $c_x$ change sign when $v_x$ changes sign
	For the bulk components from Eq. (\ref{Gb}), we find
	\beg\label{GbComponents}
	\begin{split}
	g_{z}^b=\frac{f^<_x f^>_z-f^>_x f^<_z+s^>_x s^<_z-s^<_x s^>_z}{{\cal Z}}\\
	f_{z}^b=\frac{f^<_x g^>_z-f^>_x g^<_z+g^>_x s^<_z-g^<_x s^>_z}{{\cal Z}}\\
	s_{z}^b=\frac{g_x^< f^>_z-g^>_x f^<_z+s^<_x g^>_z-s^>_x g^<_z}{\cal Z}\\
	f_{x}^b=\frac{g^<_z f^>_z-g^>_z f^<_z+s^<_x g^>_x-s^>_x g^<_x}{{\cal Z}}\\
	s_{x}^b=\frac{g^<_{x}f^>_x-g^>_x f^<_x+s^<_z g^>_z-s^>_z g^<_z}{\cal Z}\\
	g_{x}^b=\frac{s^<_x f^>_x-s^>_x f^<_x+s^<_z f^>_z - s^>_z f^<_z}{\cal Z}
	\end{split}
	\en
	where 
	\beg\label{DenomZ}
	{\cal Z}=g_z^>g_z^<+f_z^>f_z^<+s_z^>s_z^<-f_x^>f_x^<-s_x^>s_x^<-g_x^>g_x^<.
	\en
	Plugging the expressions for the functions (\ref{Anzats}) and using (\ref{axbxazbz},\ref{CoeffMore}) in Eq. (\ref{GbComponents}), we get
	\beg\label{Bulk}
	\begin{split}
		&g_z^b(\omega_n)=\frac{\omega_n}{\lambda_n}, ~f_z^b(\omega_n)=\frac{\Delta}{\lambda_n}, 
		~s_z^b(\omega_n)=\frac{M}{\lambda_n}, \\
		&s_x^b(\omega_n)=0, \quad g_x^b(\omega_n)=0, \quad f_x^b(\omega_n)=0.
	\end{split}
	\en
	Thus, we see that the method of auxiliary solution works. Note that two coefficients appearing in (\ref{axbxazbz}) and (\ref{GbComponents}) are still arbitrary and shall be fixed by the boundary conditions! 
	
	\section{General solution for clean junctions with spatially homogeneous order parameters}
	After we have determined the bulk solution from the auxiliary problems, we can now write down the general solution as the sum of bulk values and the solution of auxiliary problems multiplied by an interface dependent coefficient. For the parameters corresponding to the right-hand-side of the interface and $v_x>0$ we have
	\beg\label{Grout}
	\hat{\cal G}^r(v_x,x)=\hat{\cal G}_{\textrm{b}}^r(v_x,x)+c_r(v_x)\hat{\cal G}^r_>(v_x,x),
	\en
	where $c_r(v_x)$ is an unknown function of $v_x$ (in what follows superscripts $r/l$ refer to the functions on the right/left side of the interface). Solution with $v_x<0$ for the right-hand-side of the interface is thought as its extension to the left-hand-side:
	\beg\label{Grin}
	\hat{\cal G}^r(-|v_x|,x)=\hat{\cal G}_{\textrm{b}}^r(-|v_x|,x)+c_r(-|v_x|)\hat{\cal G}^r_<(-|v_x|,x).
	\en
	The constants $c_r(v_x)$ and $c_r(-|v_x|)$ can be found from arguments given in Ref. [\onlinecite{Yip1997}]:
	\beg\label{crvxFin}
	{
		\begin{split}
			c_r(v_x)&=\frac{2D\left\{\hat{\cal G}_{\textrm{b}}^l(0);\hat{\cal G}_<^r(0)\right\}}{\left(2-D+\frac{D}{2}\left\{\hat{\cal G}_{\textrm{b}}^r(0);\hat{\cal G}_{\textrm{b}}^l(0)\right\}\right){\cal Z}_r}, \\
			c_r(-v_x)&=
			\frac{2D\left\{\hat{\cal G}_{\textrm{b}}^l(0);\hat{\cal G}_>^r(0)\right\}}{\left(2-D+\frac{D}{2}\left\{\hat{\cal G}_{\textrm{b}}^r(0);\hat{\cal G}_{\textrm{b}}^l(0)\right\}\right){\cal Z}_r},
	\end{split}}
	\en
	where $\{\hat{X};\hat{Y}\}$ stands for the anticommutator of matrices. Similarly, for the left side, we have
	\beg\label{Glout}
	\begin{split}
		\hat{\cal G}^l(v_x,x)&=\hat{\cal G}^l_b+c_l(v_x,x)\hat{\cal G}^l_<(v_x,x),~(u<0)\\
		\hat{\cal G}^l(-|v_x|,x)&=\hat{\cal G}^l_b+c_l(-|v_x|,x)\hat{\cal G}^l_>(-|v_x|,x),~(u>0).\\
	\end{split}
	\en
	and the coefficient appearing in these equations are given by
	\beg\label{clvxFin}
	{
	\begin{split}
		c_l(v_x)&=\frac{2D\left\{\hat{\cal G}_{\textrm{b}}^r(0);\hat{\cal G}_>^l(0)\right\}}{\left(2-D+\frac{D}{2}\left\{\hat{\cal G}_{\textrm{b}}^r(0);\hat{\cal G}_{\textrm{b}}^l(0)\right\}\right){\cal Z}_l}, \\
		c_l(-v_x)&=
		\frac{2D\left\{\hat{\cal G}_{\textrm{b}}^r(0);\hat{\cal G}_<^l(0)\right\}}{\left(2-D+\frac{D}{2}\left\{\hat{\cal G}_{\textrm{b}}^r(0);\hat{\cal G}_{\textrm{b}}^l(0)\right\}\right){\cal Z}_l}.
	\end{split}}
	\en
These expressions fully describe the solution for the Eilenberger functions in clean Josephson junctions with constant order parameters and give us the quasiclassical functions which can give us profiles of order parameter and spin density wave through the self-consistency equations. 
	
The anticommutator in numerators of Eqs. (\ref{crvxFin},\ref{clvxFin}), is not diagonal and couples values of quasiclassical functions from the left side and the right side. We set for the first equation in \eqref{crvxFin}, our two constants appearing in (\ref{CoeffMore}, \ref{axbxazbz}) as
  
   \beg\label{setr1}
  c^r_{1}= \frac{c^r(M_r\Delta_l-M_l\Delta_r)\lambda^r_n}{\omega^l_n(\Delta_r^2+M_r^2)-\omega^r_n(\Delta_r \Delta_l +M_r M_l)}
  \en
and for the second equation in \eqref{crvxFin}, we change sign of $c^r_1$. Similarly, anticommutator in the numerator of the first equation in \eqref{clvxFin}, can be made diagonal by setting
 \beg\label{setl1}
c^l_{1}= \frac{c^l(M_r\Delta_l-M_l\Delta_r)\lambda^l_n}{\omega^r_n(\Delta_l^2+M_l^2)-\omega^l_n(\Delta_r \Delta_l +M_r M_l)}
\en
and changing sign of $c_l$ in the second equation in \eqref{clvxFin}. 
 This completes the proof of the method based on finding auxiliary solutions. The most interesting property of this method is that in naturally gives circuit theory rules for connectors.   
 This allows for unified treatment of superconducting junctions of arbitrary nature. 
 	
\end{appendix}

\bibliography{fesisbib}
\end{document}